\documentclass[onecolumn,prd,nofootinbib,showpacs]{revtex4}

\usepackage{textcomp}
\usepackage{mathtools, ulem, graphicx}
\usepackage[export]{adjustbox}
\usepackage{sidecap}
\usepackage{verbatim}
\usepackage{feynmp}
\usepackage{slashed}
\usepackage[caption=false]{subfig}
\usepackage{url}

\newcommand{\e}{\begin{equation*}\begin{aligned}}
\newcommand{\ee}{\end{aligned}\end{equation*}}

\newcommand{\en}{\begin{equation}\begin{aligned}}
\newcommand{\een}{\end{aligned} \end{equation}}

\newcommand{\p}{\partial}
\newcommand{\pf}[2]{\frac{\p #1}{\p #2}}
\newcommand{\f}[2]{\frac{#1}{#2}}

\newcommand{\ra}{\rangle}
\newcommand{\la}{\langle}

\newcommand{\da}{\dagger}
\newcommand{\ma}{\mathcal}

\newcommand{\Q}{\left}
\newcommand{\W}{\right}

\newcommand{\pma}{\begin{pmatrix}}
\newcommand{\epma}{\end{pmatrix}}

\newcommand{\de}{\delta}

\begin{document}

\title{Path-Integral Approach to the Scale Anomaly at Finite Temperature}
\author{Chris L. Lin}
\author{Carlos R. Ord\'{o}\~{n}ez}
\affiliation{Department of Physics, University of Houston, Houston, TX 77204-5005}

\date{\today}
\email{cllin@uh.edu}
\email{cordonez@central.uh.edu}

\begin{abstract}
We derive the relativistic thermodynamic scale equation using imaginary-time path integrals, with complex scalar field theory taken as a concrete example. We use Fujikawa's method to derive the scaling anomaly for this system using a matrix regulator. We make a general scaling argument to show how for anomalous systems, the $\beta$ function of the vacuum theory can be derived from measurement of macroscopic thermodynamic parameters. 

\end{abstract}


\maketitle

\section{Introduction}

In a series of seminal papers by Callan, Coleman, and Jackiw \cite{callan, cal}, it was noted that in general the trace of the Belifante stress-energy tensor $\theta^\mu_\mu$ for any renormalizable theory could be improved, so that classically for scale-invariant systems (systems invariant under the conformal group), 

\en \label{91}
\theta^\mu_\mu=0.
\een 

This improved tensor has a number of desirable properties over the canonical tensor (the one derived from Noether's theorem) such as having finite matrix elements in the quantum theory, and that the energy for bound states can be naturally expressed as the trace of this tensor. Shortly after these observations, it was noted that the same improvement program could be applied in the non-relativisitic case \cite{hagen}, so that for classical scale-invariant systems (systems invariant under the Schr\"{o}dinger group):

\en \label{92}
2\theta^{00}-\sum\limits_{i=1}^{3}\theta^{ii}=0,
\een  

where the $2$ results from the fact that in non-relativisitic theories time must scale as twice the power of space.\footnote{The Schr\"{o}dinger equation has only one derivative of time, and two of space, so for scale invariance time must scale as twice the power of space.} \\

Eqs. \eqref{91} and \eqref{92} fail to consider the trace anomaly. In general, the trace of the stress-energy tensor taken between bound states gives the energy of the bound state:

\en \label{93}
E_b=\int dV \la \theta^\mu_\mu \ra,
\een

which derives from the fact that the time average of the field virial is zero for bound states \cite{gate}. With slight modification Eq. \eqref{93} holds in the non-relativistic case too (see \cite{anon} for a specific example). However, it is well known that even though $\theta^\mu_\mu=0$ for a classically scale-invariant system, which would imply bound states can only have zero energy,\footnote{This is also obvious from the fact that there are no scales to even form $E_b$.} the quantization procedure can destroy this relationship. When this happens this is called a scale anomaly, and is the mechanism that allows the bound state energy to differ from zero.\\

As an example, in QCD with massless quarks (or no quarks at all), the Lagrangian is classically scale invariant so that $\theta^\mu_\mu=0$. However, through the renormalization process, a scale appears as $\Lambda_{\text{QCD}}$. In general this makes $\la \theta^\mu_\mu \ra=\ma A$, where $\ma A$ is the anomaly. The stress-energy tensor can then be further improved:

\en \label{94}
T^{\mu \nu}=\theta^{\mu \nu}+ \f{g^{\mu \nu}}{4} T^{\eta}_\eta,
\een

so that $T^{\mu \nu}$ is no longer traceless. Then

\en \label{95}
E_b=\int dV \, \Q \la T^{00}\W \ra=\int dV \, \Q \la \theta^{0 0} \W \ra+ \f{E_b}{4},
\een

which implies that $\ma A$ accounts for 1/4 of the energy of the hadron. This can explicitly be seen in the bag model where confinement of the quarks and gluons is the result of a cosmological constant term in the Lagrangian which contributes a positive energy and negative pressure $\Lambda g^{\mu \nu}$ to $\theta^{\mu \nu}$, which confines the system. Then from the tracelessness of $\theta^{\mu \nu}$, $\Lambda=\f{1}{4}T^\mu_\mu$, so that confinement accounts for 1/4 of the hadron energy \cite{gate}.\\

In this paper, we are interested in the thermal analogues of Eqns. \eqref{91} and \eqref{92}. Both of these quantities are very important in their respective areas of physics. In the nonrelativistic sector, for an ultracold dilute gas, \eqref{92} would read:

\en \label{96}
2\ma E-3P=-\f{\hbar^2}{3 m } \lambda \Q\la (\psi^\da(x) \psi(x))^2\W \ra.
\een

The RHS is known as the Tan contact, and is extremely important in atomic physics. In terms of it, Tan derived a set of universal relations \cite{tan1,tan2,tan3} that govern many relationships between the thermodynamics variables of the system and the behavior of the large momentum tails of correlation functions. These relationships hold even in the strongly interacting regime where perturbation theory becomes inadequate \cite{braught}. A field theoretic explanation of Tan's result was later developed in terms of the operator product expansion \cite{brat}. \\

In QCD, the analog would be \cite{notes}:

\en \label{97}
\ma E-3P=\sum\limits_{i=1}^{n_f} m_i \Q\la \bar{\psi}_i\psi_i\W \ra+\f{2}{g}\beta(g)\f{1}{4}\Q\la F^a_{\mu \nu}F^{\mu \nu a}\W \ra.
\een

In the low temperature regime where the coupling $g$ is strong, the trace anomaly of the RHS is calculated by calculating the LHS of Eq. \eqref{97} using a lattice action. The goal is to calculate the QCD equation of state $P=P(T,\mu,V)$ rather than the anomaly itself. However, for technical reasons \cite{lattice}, $\ma E-3P$ is important as an intermediate step in lattice QCD for calculating $P(T,\mu,V)$, where it is given by:

\en \label{98}
\ma A=\ma E-3P=-\f{T}{V}\f{d\ln Z}{d\ln a},
\een 

and plugging into Eq. \eqref{97} gives after using thermodynamic identities:

\en \label{99}
\f{\p }{\p \ln T}\Q(\f{P}{T^4} \W)=\f{\ma A}{T^4},
\een

which can then be integrated to get $P(T,\mu,V)$. $a$ is the lattice spacing and $Z$ is the partition function with lattice action. \\

In this paper, following the approach initiated in \cite{Ord, vir, lin} for non-relativisitic systems, we provide a continuum/non-lattice path-integral approach to deriving the thermodynamic trace equation $\ma E-3P$, where anomalies naturally appear as a result of a change of variables of the path-integral measure, the thermal analog of Fujikawa's method. This is in contrast to an operator approach, where one takes the thermal quantum statistical expectation values of both sides of Eqns. like \eqref{91} and \eqref{92}, and identifying $\Q \la T^{00} \W \ra=\ma E$ and $\Q \la T^{ii} \W \ra=\ma P_H$, where $\ma P_H$ is the hydrodynamic pressure \cite{pressure}. Within this path-integral approach, no reference needs to be made about improvement of the stress-energy tensor, or the validity of equating the hydrodynamic pressure $\ma P_H$ with the thermodynamic pressure $P$ derived from the grand partition function, which is nontrivial, especially in the presence of anomalies \cite{weert, cern}. For concreteness, we will take as our system a complex scalar field theory, but the results can be extended for other systems. The Lagrangian is given by

\en \label{1}
\ma L=\p^\mu\phi^\da \p_\mu \phi-m^2\phi^\da \phi-\f{\lambda}{4}(\phi^\da \phi)^2
\een

and has a $U(1)$ symmetry 

\en \label{2}
\phi &\rightarrow e^{i\theta} \phi, \\
\phi^\da &\rightarrow e^{-i\theta} \phi^\da,
\een

leading to a conserved charge:

\en \label{3}
j_0&= i\phi^\da \overset{\leftrightarrow}{\partial_0} \phi, \\
Q&=i\int d^3x \, \phi^\da \overset{\leftrightarrow}{\partial_0} \phi .
\een

Under scale transformation:

\en \label{inserthere}
x'^\mu&=e^\rho x^\mu ,\\
\phi'(x')&=e^{-\rho}\phi(x), \\
\phi'^\da(x')&=e^{-\rho}\phi^\da(x).
\een

\section{Thermodynamic Dilation Equation}

For a homogeneous system the grand potential $\Omega=\Omega(\beta,\mu,V)$ in the large volume limit equals $-PV$, so that the partition function is $Z=e^{-\beta \Omega}=e^{\beta PV}$, and can be expressed via a path integral:

\en \label{4}
Z=e^{\beta PV}=\sum\limits_i \la i |e^{-\beta (H-\mu Q)}| i \ra= \int [d\phi][d\phi^*] e^{-S_E+\mu \int_0^\beta \int_V d^3xd\tau \, j_0},
\een

with\footnote{Due to the dependence of $j_0$ on conjugate momenta, when integrating out conjugate momenta to pass into the Lagrangian formulation of the path integral, $\ma L_E$ acquires an additional $\mu^2\phi^*\phi$ term: see \cite{kap}.}

\en \label{5}
S_E&=\int_0^\beta \int_V d^3xd\tau \, \Q( \p_\mu\phi^* \p_\mu \phi+(m^2-\mu^2)\phi^* \phi+\f{\lambda}{4}(\phi^* \phi)^2\W),\\
j_0&=-\phi^* \overset{\leftrightarrow}{\partial_\tau} \phi . 
\een

Now consider an infinitesimal ``relativistic thermodynamic scaling'' 

\en \label{6}
\beta'&=e^{\rho} \beta=\beta+\rho \beta=\beta+\delta \beta, \\
L_i'&=e^{\rho} L_i=L_i+\rho L_i=L_i+\delta L_i, \\
\mu'&=\mu.
\een

where $L_i$ is the length of the box in the $i$ direction and $\rho$ is a dimensionless infinitesimal parameter. \\

In the large volume limit it is assumed that $P(\beta,\mu,V)=P(\beta,\mu)$,\footnote{This can be shown via cluster decomposition: e.g., see \cite{vir}.} so under the transformation of Eq. \eqref{6}:

\en \label{7}
\delta (\beta PV)&=(\delta \beta) PV+\beta (\delta P) V+\beta P (\delta V)    \\
&=\rho \Q(\beta PV+\beta \Q(\f{\p P}{\p\beta}\beta \W)V+\beta P (3V)   \W).
\een

Now using the identity $\beta V \f{\p P}{\p \beta}=-PV-E+\mu Q$, we get

\en \label{8}
\delta (\beta PV)=\rho \Q(-\beta E+\beta P (3V)+\beta \mu Q \W),  
\een

and therefore

\en \label{9}
\delta \Q(e^{\beta PV}\W)=\delta \Q(\beta PV\W) e^{\beta PV}=\rho \beta \Q(-E+3PV+\mu Q \W) e^{\beta PV}.
\een

Eq. \eqref{9} represents the effect of the scaling in Eq. \eqref{6} on the LHS of Eq. \eqref{4}. Now we analyze the effect of this scaling to the RHS of Eq. \eqref{4}, the path-integral part, from which anomalies originate, and eventually equate the two expressions. \\

The scaling in Eq. \eqref{6} represents a dilation of the system:

\en \label{10}
x'^\mu&=e^{\rho} x^\mu ,\\
\phi'(x')&=e^{-\rho}\phi (x),\\
\phi'^*(x')&=e^{-\rho}\phi^*(x).
\een

The dilated system has 

\en \label{11}
e^{\beta' P'V'}=\int [d\phi'][d\phi'^*] e^{-S'_E+\mu \int_0^{\beta'} \int_{V'} d^Dx'd\tau' \, j'_0},
\een 

where 

\en \label{12}
S'_E&=\int_0^{e^{\rho} \beta} \int_{e^{\rho} V} d^3x'd\tau' \, \Q( \p'_\mu\phi'^* \p'_\mu \phi'+(m^2-\mu^2)\phi'^* \phi'+\f{\lambda}{4}(\phi'^* \phi')^2\W),\\
\mu \int_0^{\beta'} \int_{V'} d^3x'd\tau' \, j'_0 &=\mu \int_0^{e^{\rho} \beta} \int_{e^{\rho} V} d^3x'd\tau' \, \Q(-\phi'^* \overset{\leftrightarrow}{\partial'_\tau} \phi' \W).
\een

To compare to the undilated system, we ``pull back'' to unprimed variables by substituting Eq. \eqref{10} into Eq. \eqref{11} and Eq. \eqref{12}. Eq. \eqref{12} becomes:

\en \label{13}
S'_E&=\int_0^{e^{\rho} \beta} \int_{e^{\rho} V} d^3x'd\tau' \, \Q( \p'_\mu\phi'^* \p'_\mu \phi'+(m^2-\mu^2)\phi'^* \phi'+\f{\lambda}{4}(\phi'^* \phi')^2\W)\\
&=\int_0^{\beta} \int_{V} e^{4\rho} d^3x d\tau \, \Q(e^{-2\rho} \f{\p \phi^*}{ \p \Q(e^{\rho} x_\mu\W)} \f{\p \phi}{ \p \Q(e^{\rho} x_\mu\W)} +(m^2-\mu^2)e^{-2\rho}\phi^* \phi+\f{\lambda}{4}(e^{-2\rho}\phi^* \phi)^2\W)\\
&=S_E+2\rho\int_0^{\beta} \int_{V} d^3x d\tau \,(m^2-\mu^2) \phi^* \phi.
\een

Similarly:
\en \label{14}
\mu \int_0^{\beta'} \int_{V'} d^3x'd\tau' \, j'_0&=\mu \int_0^{e^{\rho} \beta} \int_{e^{\rho} V} d^3x'd\tau' \, \Q(-\phi'^* \overset{\leftrightarrow}{\partial'_\tau} \phi' \W)\\
&=\mu \int_0^{\beta} \int_{V} d^3xd\tau \, j_0+\rho \mu \int_0^{\beta} \int_{V} d^3xd\tau \, j_0.
\een

Plugging in these expressions into Eq. \eqref{11}:

\en \label{15}
e^{\beta' P'V'}=\int J[d\phi][d\phi^*] e^{-S_E+\mu \int_0^{\beta} \int_{V} d^3xd\tau \, j_0-2\rho\int_0^{\beta} \int_{V} d^3x d\tau \,(m^2-\mu^2) \phi^* \phi +
\rho \mu \int_0^{\beta} \int_{V} d^3xd\tau \, j_0},
\een

where $J$ is the Jacobian of the transformation $(\phi',\phi'^*) \rightarrow (\phi,\phi^*)$. Expressing $J=1-\rho A$ and using Eq. \eqref{9}: \footnote{
$\la F(\phi,\phi^\da)\ra \equiv \f{1}{Z}\int [d\phi][d\phi^*]  F(\phi,\phi^*) e^{-S_E+\mu \int_0^{\beta} \int_{V} d^3xd\tau \, j_0}$. }

\en \label{16}
\delta\Q(e^{\beta PV}\W)=\rho \beta \Q(-E+3PV+\mu Q \W) e^{\beta PV}=\rho \Q( -A-2 \Q\la \int_0^{\beta} \int_{V} d^3x d\tau \,(m^2-\mu^2)\phi^\da \phi \W\ra
+\Q\la  \mu \int_0^{\beta} \int_{V} d^3xd\tau \, j_0 \W \ra \W)e^{\beta PV}.
\een
 
The chemical potential terms drop out on both sides\footnote{Using the identity $Q=\f{\p P}{\p \mu}$ and Eq. \eqref{4}, $Q=\f{\p P}{\p \mu}=\Q \la j_0 \W \ra + 2\mu \Q \la \phi^\da \phi \W \ra$.} and we get:

\en \label{17}
\ma E-3P=2m^2 \Q\la \phi^\da \phi \W\ra+\ma A,
\een

where 

\en \label{18}
J&=\Q[\f{\p \phi'\p \phi'^*}{\p \phi \p \phi^*}\W]=e^{\text{Tr} \log\Q(I_2 \Q( \delta^4(x-y)+\rho(-1-x_\mu\p_\mu)\delta^4(x-y)\W)\W)}\\
&=e^{\rho \int d^4x\,\text{tr}\Q[(-1-x_\mu\p_\mu)\delta^4(x-y) I_2\W]\big|_{x=y}}\\
&=1+\rho \int d^4x\, \text{tr}\Q[(-1-x_\mu\p_\mu)\delta^4(x-y) I_2\W]\big |_{x=y},
\een

so that

\en \label{19}
\ma A=\text{tr}\Q[(1+x_\mu\p_\mu)\delta^4(x-y) I_2\W]\big |_{x=y}.
\een

$I_2$ is the two dimensional identity matrix which results from having two fields, $\phi$ and $\phi^*$.\footnote{Note that $\text{Tr}$ in Eq. \eqref{18} refers to both discrete ($2 \times 2$) and continuous variables, whereas $\text{tr}$ in Eq. \eqref{19} refers to only ($2\times 2$).} $\ma A=\f{A}{\beta V}$ is the anomaly, a divergent quantity that requires regularization.

\section{Fujikawa Calculation}
In Euclidean space, $\ma L_E=\p_\mu\phi^\da \p_\mu \phi
+m^2 \phi^\da \phi+\f{\lambda}{4}(\phi^\da \phi)^2$. A saddle point expansion about a constant classical background $\phi$ produces the quadratic piece $\ma L_2$:

\en \label{20}
\ma L_2&=\f{1}{2}\pma  \eta^\da & \eta   \epma
\pma
-\p^2+m^2+\lambda \phi^* \phi  &  \f{\lambda}{2} \phi \phi \\
\f{\lambda}{2} \phi^*\phi^*  & -\p^2+m^2+\lambda \phi^* \phi
\epma
\pma \eta \\ \eta^\da\epma \\
&\equiv
\f{1}{2}\pma  \eta^\da & \eta   \epma
\pma
-\p^2+C  &  \f{\lambda}{2} \phi \phi \\
\f{\lambda}{2} \phi^*\phi^*  & -\p^2+C
\epma
\pma \eta \\ \eta^\da\epma\\
& \equiv
\f{1}{2}\pma  \eta^\da & \eta   \epma
M
\pma \eta \\ \eta^\da\epma, 
\een

where $C=m^2+\lambda \phi^* \phi$, $\eta$ is the fluctuating field around $\phi$, and $M$ is a Hermitian matrix. Following Fujikawa \cite{fujikawa}, we use $M$, the bilinear matrix, as the Hermitian matrix that goes in our regulator\footnote{e.g., for the chiral anomaly with $\ma L=\bar{\psi}i\slashed{D}\psi$, the matrix $i\slashed{D}$ is to be used as the argument of the regulator. $M$, the quadratic piece of the quantum action, naturally captures the 1-loop effects of interactions which are responsible for anomalies.}. Choose a regulator of the form $R=R\Q(\f{M}{\Lambda^2}\W)$ with the property that $R(0)=I_2$. The expression to be regulated is:

\en \label{21}
\ma A=\text{tr}\pma
\theta \de(x-y) & 0 \\ 0 &  \theta\de(x-y)
\epma \Bigg |_{x=y}
\een

where $\theta=1+x_\mu\p_\mu$, so that 

\en \label{22}
\ma A_R=\text{tr}\Q[R\Q(\f{M}{\Lambda^2}\W)\theta\delta(x-y)I_2\W]\Bigg |_{x=y}.
\een

This expression equals:

\en \label{23}
\ma A_R&=\int \f{d^4k}{(2\pi)^4}\text{tr}\,R\pma
\f{-\p^2+C}{\Lambda^2}  &  \f{\lambda \phi \phi}{2\Lambda^2} \\
\f{\lambda \phi^*\phi^*}{2\Lambda^2}  & \f{-\p^2+C}{\Lambda^2}
\epma \theta e^{-ik(x-y)}\Bigg |_{x=y} \\
&=
\int \f{d^4k}{(2\pi)^4}\text{tr}\,R\pma
\f{k^2+C}{\Lambda^2}  &  \f{\lambda \phi \phi}{2\Lambda^2} \\
\f{\lambda \phi^*\phi^*}{2\Lambda^2}  & \f{k^2+C}{\Lambda^2}
\epma (1-ix_\mu k_\mu) \\
&=\Lambda^4\int \f{d^4k}{(2\pi)^4}\text{tr}\,R\pma
k^2+\f{C}{\Lambda^2}  &  \f{\lambda \phi \phi}{2\Lambda^2} \\
\f{\lambda \phi^*\phi^*}{2\Lambda^2}  & k^2+\f{C}{\Lambda^2}
\epma (1-i\Lambda x_\mu k_\mu) \\
&=
\Lambda^4\int \f{d^4k}{(2\pi)^4}\text{tr}\,R\pma
k^2+\f{C}{\Lambda^2}  &  \f{\lambda \phi \phi}{2\Lambda^2} \\
\f{\lambda \phi^*\phi^*}{2\Lambda^2}  & k^2+\f{C}{\Lambda^2}
\epma, 
\een

where the $k_\mu$ term is odd so vanishes over the integral when multiplied by the even function $R(-k)=R(k)=f(k^2)$. Next we define:

\en \label{24}
D&=k^2 I_2, \\
B&=\f{1}{\Lambda^2}\pma
C  &  \f{\lambda \phi \phi}{2} \\
\f{\lambda \phi^*\phi^*}{2}  & C
\epma, 
\een

so that the equation can be written succinctly:

\en \label{25}
\ma A_R=\Lambda^4\int \f{d^4k}{(2\pi)^4} \text{tr}\,R(D+B).
\een

We then Taylor expand about $D$ (note that $[D,B]=0$ so the Taylor expansion is valid):

\en \label{26}
\ma A_R=\Lambda^4\int \f{d^4k}{(2\pi)^4} \text{tr}\, \Q(R(D)+R'(D)B+\f{1}{2}R''(D)B^2+...  \W).
\een 

The first term is the same as in the non-interacting case, which is taken to be anomaly free \cite{ume}, so we neglect it. The second term can be absorbed by a mass counterterm. Terms higher order than the third term fall faster than $\f{1}{\Lambda^4}$ so the $\Lambda^4$ prefactor in Eq. \eqref{26} cannot keep them from going to zero. Only the 3rd term is independent of the cutoff. Therefore:

\en \label{27}
\ma A_R&=\Lambda^4\int \f{d^4k}{(2\pi)^4} \f{1}{2}\text{tr}\Q(R''(D)B^2\W) \\
&=\Lambda^4\int \f{k^2 dk^2}{16\pi^2} \f{1}{2}\text{tr}\Q(R''(D)B^2\W),
\een 

where the solid angle $\Omega=2\pi^2$ was used. Now 

\en \label{insert}
B^2=\f{1}{\Lambda^2}\pma
C^2+\f{\lambda^2 (\phi^* \phi)^2}{4}&\lambda C \phi \phi\\
\lambda C \phi^* \phi^* &C^2+\f{\lambda^2(\phi^* \phi)^2}{4}
\epma\equiv
\f{1}{\Lambda^2}\pma
B_1 & B_2 \\
B^*_2 & B_1 
\epma,
\een

and since $R(D)$ is diagonal, we can define:

\en \label{28}
R(D)=f(k^2) I_2.
\een

Note that the derivative in Eq. \eqref{27} is w.r.t. $k^2$. Therefore:

\en \label{29}
\ma A_R&=\Lambda^4\int \f{k^2 dk^2}{16\pi^2} \f{1}{2}\text{tr}\Q( R''(D)B^2\W)\\
&=B_1 \int \f{k^2 dk^2}{16\pi^2} f''(k^2),
\een

where we have safely taken $\Lambda \rightarrow \infty$. Integrating by parts:

\en \label{30}
\ma A_R &= \f{B_1}{16\pi^2} \Q[k^2f'(k^2)\W]\Big|^{\infty}_{0}-\f{B_1}{16\pi^2} \int dk^2 f'(k^2) \\
&=\f{B_1}{16\pi^2} \Q[k^2f'(k^2)\W]\Big|^{\infty}_{0}-\f{B_1}{16\pi^2} f(k^2)\Big|^{\infty}_{0}\\
&=\f{B_1}{16\pi^2},
\een

where we require 

\en \label{31}
f(0)&=1 \\ 
f(\infty)&=0 \\
\Q[k^2 f'(k^2)\W]\big |^\infty_0&=0, 
\een

which are the same conditions on the regulator for the chiral case \cite{Weinberg}. \\

Plugging in $B_1$ from Eq. \eqref{insert} into Eq. \eqref{30}, we get:

\en
\ma A_R=\f{C^2+\f{\lambda^2 (\phi^* \phi)^2}{4}}{16\pi^2}=\f{5\lambda^2 (\phi^* \phi )^2}{64 \pi^2}+\f{m^4}{16\pi^2}+\f{\lambda m^2 (\phi^* \phi)}{8\pi^2}.
\een

The second term is independent of the coupling, and since the free theory is taken to be non-anomalous, we can subtract it. The third term can be absorbed into the mass term of Eq. \eqref{17}, leaving only the 1st term as the anomaly \cite{shiz}. Therefore

\en \label{insert2}
\ma E-3P=\f{5\lambda^2}{64 \pi^2}  \Q \la (\phi^\da \phi )^2 \W \ra.
\een

Note that the anomaly $\ma A_R$ occurs inside the path integral, and $\f{1}{Z}\int [d\phi d\phi^* ] f(\phi,\phi^*)e^{-S_E+...}=\la f(\phi,\phi^\da)\ra$, so that in Eq. \eqref{insert2} there are expectation values. This replacement is valid up to 1-loop \cite{shiz}.

\section{Dimensional Analysis for Relativistic Systems} \label{dimension}

In relativistic theories we set $\hbar=c=k_B=1$. The units for all quantities can then be written as $\hbar^ic^jk_B^kL^\ell$, where $L$ is a variable in the problem with units of length. Suppose the system has microscopic parameters $g_k$, which can be coupling constants or dimensionally transmuted quantities. We define $[g_k]=\ell$ as the power of $L$ when $g_k$ is written in units of $\hbar^ic^jk_B^kL^\ell$. So for example $[m]=[E]=-1$. The grand potential $\Omega=\Omega(\beta, \mu_i, V, g_i)$ has $[\Omega]=-1$ and can be written as:

\en \label{32}
\Omega(\beta, z_i, V, g_i)=V \beta^{-1-D} f(z_i, g_i \beta^{-[g_i]}),
\een

where $f(z_i, g_i \beta^{-[g_i]})$ is a dimensionless function of dimensionless variables, $z_i$ is the fugacity corresponding to $\mu_i$ ($z_i=e^{\beta \mu_i}$), and $D$ is the number of spatial dimensions.\footnote{For example, if the coupling $g_1$ has dimensions of length, the corresponding dimensionless variable is $g_1 \beta^{-1}=g_1 T$ which is dimensionless. If the coupling $g_2$ as dimensions of energy, $g_2 \beta^{-(-1)}=g_2\beta=\f{g_2}{T}$.} $\Omega$ has this form because $\beta$ and $\mu_i$ don't depend on the absolute size of the system (they are intensive variables). If one doubles the system keeping $\beta$ and $\mu_i$ constant, then $\Omega$, being an extensive quantity, should double. So $\Omega$ must be proportional to V.\footnote{$\Omega=-PV$, so Eq. \eqref{32} is consistent with the statement that $P(\beta,\mu,V)=P(\beta,\mu)$.} To make up for the remaining dimension ($[\Omega]=-1$), we are free to pull out one of the dimensionful arguments of $\Omega$, and the rest of the arguments must be ratios with the argument we pulled out. We will pull out $\beta$. This is equivalent to choosing our scale as $\beta$ and measuring all other quantities in units of $\beta$.\\

Take the derivative of Eq. \eqref{32} w.r.t. to $\beta$ at constant fugacity $z_i$ and volume $V$, and multiply times $\beta$:

\en \label{33}
\begin{split}
\left.\beta \pf{\Omega}{\beta}\right|_{z_i,V}&=\Q(-1-D\W)\Omega+V \beta^{-1-D} \beta \left.\pf{ f(z_i,g_i \beta^{-[g_i]})}{\beta}\right|_{z_i}\\
&=\Q(-1-D\W)\Omega+V \beta^{-1-D} \beta \Q[\sum\limits_{k}\f{-[g_k]g_k}{\beta} \left.\pf{ f(z_i,g_i \beta^{-[g_i]})}{g_k}\W]\right|_{z_i}\\
&=\Q(-1-D\W)\Omega-\sum\limits_{k}[g_k]g_k \pf{\Omega}{g_k}.
\end{split}
\een

Now, we use the thermodynamic identity $E=\left. \pf{(\beta \Omega)}{\beta}\right|_{z_i,V}=\Omega+\left .\beta \pf{ \Omega}{\beta}\right|_{z_i,V}$. 
\en \label{34}
\begin{split}
E-DPV&=\Q(\Omega+\left .\beta \pf{ \Omega}{\beta}\right|_{z_i,V} \W)-DPV\\
&=\Q(\Omega+\Q(-1-D\W)\Omega-\sum\limits_{k} [g_k]g_k \pf{\Omega}{g_k} \W)-DPV\\
&=-\Q(P+\Q(-1-D\W)P-\sum\limits_{k} [g_k]g_k \pf{P}{g_k} \W)V-DPV\\
&=\sum\limits_{k}[g_k]g_k \pf{P}{g_k}V \\
\ma E-DP&=\sum\limits_{k} [g_k] g_k \pf{P}{g_k}. 
\end{split}
\een

where the derivatives are at constant $\beta$, $\mu$, and $V$. \\

\section{$\beta$ Function}
For a system that develops a microscopic scale $M$ through dimensional transmutation via renormalization of the coupling constant:

\en \label{36}
\ma E-DP=[M] M \f{d\lambda}{dM} \pf{P}{\lambda}=-M \f{d\lambda}{dM} \pf{P}{\lambda}=-\beta(\lambda)\pf{P}{\lambda}=\beta(\lambda) \Q \la \f{\p \ma H_I}{\p \lambda} \W\ra,
\een

since $\f{\p P}{\p \lambda}=\f{1}{\beta V} \f{\p}{\p\lambda}\ln \Q\{\int [d\phi][d\phi^*] e^{-S_E+\mu \int d^Dxd\tau j^o} \W\}$ pulls down the interaction term in the path integral, creating a thermal average. \\

Comparison of Eq. \eqref{1}, Eq. \eqref{17}, Eq. \eqref{insert2}, and Eq. \eqref{36} gives:

\en \label{37}
\beta(\lambda)=\f{5\lambda^2}{16\pi^2},
\een

as 

\en \label{38}
\ma E-3P=\f{5\lambda^2}{64\pi^2}\Q \la (\phi^\da \phi )^2 \W \ra=\beta(\lambda) \Q\la \f{(\phi^\da \phi)^2}{4} \W\ra
\een

would give Eq. \eqref{37}.\\

\begin{center}
\begin{figure}
\includegraphics[trim = 0mm 0cm 0mm 0mm, clip, width=8cm]{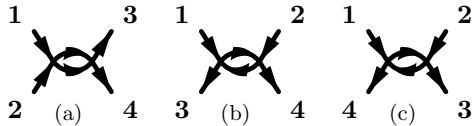}
\caption{Diagrams contributing to the $\beta$ function for complex scalar field theory. 1 and 2 refer to incoming particles, 3 and 4 to outgoing particles.}
\label{fig1}
\end{figure}
\end{center}

The $\beta$ function of Eq. \eqref{37} can be gotten from setting $e=0$ for the charge $e$ in the calculation for the four-scalar vertex in scalar electrodynamics \cite{sred}. A diagrammatic calculation requires the identification of 3 diagrams (see Fig \ref{fig1}). Diagram (a) contains a symmetry factor of 1/2 due to the swapping of internal propagators. Modulo the symmetry factor, each diagram contributes the same amount to the $\beta$ function, giving $1/2+1+1=5(1/2)$, or the first diagram's contribution multiplied by $5$. The matrix $M$ used for regularization automatically mixes the interactions, giving the factor of 5. Using the definition of the beta function $M \f{d\lambda }{dM}=\f{5\lambda^2}{16\pi^2}$ and setting the renormalization scale $M=T$, one can solve the differential equation for the coupling $\lambda(T)=\f{16\pi^2}{5\ln\Q(\f{\Lambda}{T}\W)}$, where $\Lambda$ is the Landau pole. As $\f{T}{\Lambda} \rightarrow 0$ the coupling is small and the system behaves like a gas of noninteracting bosons, while as $T \rightarrow \Lambda$ the coupling blows up and perturbation theory fails.  \\

\section{Conclusions}

In this paper we have extended to relativistic systems the path-integral approach to the study of quantum anomalies for many-body systems initiated in \cite{Ord, lin,vir}.  
A notable difference is that in the relativisitic case we have a very wide class of regulators characterized by the function $f(k^2)$ of Eq. \eqref{28}, which other than satisfying Eq. \eqref{31}, is of a very general nature. An interesting result of this paper is the extraction of the leading order result for the beta function for complex fields, Eq. \eqref{37}, obtained here by comparing Eqs. \eqref{1}, \eqref{17}, \eqref{insert2} and \eqref{36}, without resorting to graphical methods \cite{wein, sred}. This result gives further support to the importance of Fujikawa's approach in the description of quantum anomalies for systems at finite temperature and density. We are currently pursuing further studies and extensions of this method, as well as applications to other systems with classical scale symmetry.

\begin{acknowledgements}
This work was supported in part by the U.S. Army Research Office Grant No. W911NF-15-1-0445.
\end{acknowledgements}

\end{document}